\documentclass[aps,prb,twocolumn,superscriptaddress,showpacs,showkeys]{revtex4-2}
\usepackage[pdftex]{graphicx}
\usepackage{color}
\usepackage{amsmath}
\usepackage{multirow}
\usepackage{placeins}
\usepackage{filecontents}

\usepackage{epstopdf}
\usepackage{changes}

\usepackage{xr}
\begin{document}

\newcommand{\ALBA}[0]{{
ALBA Synchrotron Light Source, Carretera BP 1413 km 3.3, 08290 Cerdanyola del Vall{\`e}s, Spain}}
\newcommand{\CFM}[0]{{
Centro de F\'{\i}sica de Materiales CSIC/UPV-EHU-Materials Physics Center, E-20018 San Sebasti\'an, Spain}}
\newcommand{\UPVPMA}[0]{{
Departamento de Pol\'imeros y Materiales Avanzados: F\'isica, Qu\'imica y Tecnolog\'ia,
Universidad del Pa\'is Vasco UPV-EHU, 20018 San Sebasti\'an, Spain}}
\newcommand{\DIPC}[0]{{
Donostia International Physics Center, E-20018 Donostia-San Sebasti\'an, Spain}}
\newcommand{\UPVFA}[0]{{
Departamento de F\'isica Aplicada I,
Universidad del Pa\'is Vasco UPV/EHU,  E-20018 San Sebasti\'an, Spain}}
\newcommand{\IOMCNR}[0]{{
IOM-CNR, Laboratorio TASC, Strada Statale 14 Km 163.5, I-34149 Trieste, Italy}}
\newcommand{\Elettra}[0]{{
Elettra - Sincrotrone Trieste S.C.p.A., Strada Statale 14 Km 163.5, I-34149 Trieste, Italy}}
\newcommand{\TUDD}[0]{{
Institut f\"ur Festk\"orper- und Materialphysik, Technische Universit\"at
Dresden, D-01062 Dresden, Germany}}
\newcommand{\Soleil}[0]{{
Synchrotron SOLEIL, CNRS-CEA, L'Orme des Merisiers, Saint-Aubin-BP48, 91192 Gif-sur-Yvette, France}}
\newcommand{\ISMTrieste}[0]{{
Instituto di Struttura della Materia, Consiglio Nazionale delle Ricerche, 34149 Trieste, Italy}}

\author{M. Blanco-Rey}
\affiliation{\UPVPMA}
\affiliation{\DIPC}
\author{R. Castrillo-Bodero}
\affiliation{\CFM}
\author{K. Ali}
\affiliation{\DIPC}
\affiliation{\CFM}
\author{P. Gargiani}
\affiliation{\ALBA}
\author{F. Bertran}
\affiliation{\Soleil}
\author{P.M. Sheverdyaeva}
\affiliation{\ISMTrieste}
\author{J.E. Ortega}
\affiliation{\UPVFA}
\affiliation{\CFM}
\affiliation{\DIPC}
\author{L. Fernandez}
\affiliation{\UPVFA}
\affiliation{\CFM}
\author{F. Schiller}
\affiliation{\CFM}
\affiliation{\DIPC}
\email{frederikmichael.schiller@ehu.es}

\title{Valence state determines the band magnetocrystalline anisotropy in 2D rare-earth/noble-metal compounds}

\pagebreak

\date{\today}

\begin{abstract}
In intermetallic compounds with zero-orbital momentum ($L=0$) 
the magnetic anisotropy and the electronic band structure are 
interconnected. Here, we investigate this connection on divalent 
Eu and trivalent Gd intermetallic compounds. 
We find by X-ray magnetic circular dichroism an out-of-plane easy 
magetization axis in 2D atom-thick EuAu$_2$.
Angle-resolved photoemission and density-functional theory prove that this is
due to strong $f-d$ band hybridization and Eu$^{2+}$ valence. In contrast, the
easy in-plane magnetization of the structurally-equivalent GdAu$_2$ is ruled by
spin-orbit-split $d$-bands, notably Weyl nodal lines, occupied in the Gd$^{3+}$ state.
Regardless of the $L$ value, 
we predict a similar itinerant electron contribution
to the anisotropy of analogous compounds.
\end{abstract}

\newpage
\maketitle

\section{Introduction}

The spin-orbit coupling (SOC) is responsible for the
splitting of a $4f^n$ orbital into multiplet states $|J J_z\rangle$
defined by the total angular momentum $J$ and its projection $J_z$ on the
magnetization direction. If these levels are accessible
by the crystal field energies, quenching of $\langle J_z \rangle$ can occur.
The orientation dependence of this mixing of states is at the origin of the strong
magnetocrystalline anisotropy of systems based in rare earths (RE)
\cite{bib:buschowbook,bib:skomski09}.
In recent years, magnets in the zero-dimensional (0D) limit have
been successfully realized using REs, either as isolated
ad-atoms~\cite{bib:donati14,bib:donati16,bib:singha17,bib:natterer18,bib:donati20} or as centers in organic 
molecules~\cite{bib:ishikawa05,bib:rinehart11,bib:martinez12,bib:mannini14,bib:heinrich15,bib:chiesa20,bib:briganti21}.
Prior to these achievements it was known that
the $4f$ electron shell confers to bulk intermetallic RE compounds uniaxial anisotropy and
high Curie temperatures $T_C$ (and consequently large coercive fields)~\cite{bib:givord85},
where the magnetic coupling follows a RKKY mechanism~\cite{bib:buschow79}.
In two-dimensional (2D) rare-earth/noble-metal (RE/NM) surface compounds large $T_C$ values
are retained~\cite{bib:corso10,bib:corso10b,bib:ormaza13,bib:fernandez14,bib:ormaza16,bib:fernandez20}.
Anisotropy is a requisite for stable
long-range magnetic ordering in 2D~\cite{bib:mermin66,bib:berezinskii71,bib:kosterlitz73}.
Magneto-optic Kerr effect (MOKE) and
X-ray magnetic circular dichroism (XMCD) studies reveal that the
magnetocrystalline anisotropy in these compounds does not follow a common trend.
For example, the HoAu$_2$ monolayer (ML) grown on Au(111) is strongly 
anisotropic with out-of-plane (OOP) easy axis,
whereas the GdAu$_2$ one is more easily magnetized in-plane (IP).
The HoAu$_2$ case can be understood
in terms of the $^5H_8$ state given by Hund's rule for 4$f^{10}$ (Ho$^{3+}$),
since the oblate shape of the orbital and the surface charge distribution
favour an OOP easy axis~\cite{bib:rinehart11,bib:fernandez20}. However,
explaining the GdAu$_2$ anisotropy demands an extension of the model.
In fact, the 4$f$ shell of Gd$^{3+}$ is half-filled,
i.e. the total orbital quantum number is $L=0$ ($^8S_{7/2}$ ground state),
and the hybridization with the surface bands is negligible,
and therefore the anisotropy must have a different origin.

At half-filling $4f^7$, classical dipolar interactions tend to dominate
the anisotropy in bulk Gd compounds~\cite{bib:geldart89,bib:rotter03}, yet
the magnetocrystalline contribution is not fully quenched, as
observed in metallic $hcp$ Gd \cite{bib:franse80}.
The latter results from the spin polarization of the $5d$ conduction electrons,
which are also subject to SOC \cite{bib:wietstruk11,bib:colarieti03}.
This leads to sizable magnetocrystalline 
anisotropy energy (MAE) values and non-zero orbital momenta \cite{bib:jang16}.

In this work, we show that the magnetocrystalline anisotropy of RE-NM$_2$ MLs
is not defined by the $4f$ single-orbital anisotropy only, but there exists 
an additional term originated at the itinerant electrons. 
Since all the compounds of the RE-NM$_2$ family display similar band dispersion features,
the calculations presented here for $L=0$ systems, namely EuAu$_2$ and GdAu$_2$, predict that
the itinerant electrons contribute to the MAE with $\approx 1$\,meV in general and that the
RE valence state determines whether this contribution favours an OOP or IP easy axis of magnetization.
This means that, for RE materials with a large 4$f$ single orbital anisotropy, 
the total anisotropy will not depend on the band dispersion. 
For other RE metals, however, the band dispersion may define the magnetic anisotropy. 
We show that Eu in the EuAu$_2$ ML on Au(111) behaves as a divalent species and thus it is
nominally in a $^8S_{7/2}$ ground state, as Gd$^{3+}$ in the GdAu$_2$ ML.
However, an EuAu$_2$ ML presents an OOP easy axis of magnetization and  $\mathrm{MAE}=1.6$\,meV,
in contrast to the IP easy axis observed in GdAu$_2$.
The IP easy axis of GdAu$_2$ is explained in
terms of the SOC lifting degeneracies in dispersive valence bands with Gd($d$) character.
On the other hand, in divalent EuAu$_2$ the Eu($d$) states are
unoccupied and do not contribute.
Instead, the anisotropy is caused by the strong $f-d$ band hybridization.

\section{Methods}
\subsection{Experimental Methods}

Sample preparation of EuAu$_2$ has been carried out by thermal deposition of
Eu onto a clean Au(111) single crystalline surface.
The formation of the monolayer is achieved when a complete layer 
with moir\'e is observed in Scanning Tunneling Microscopy (STM), 
or the surface state emission of the Au(111) Shockley state in angle-resolved
photoemission spectroscopy (ARPES)~\cite{bib:reinert01,bib:nicolay01}
has completely vanished. 
The substrate temperature is around 675\,K for GdAu$_2$, 
whereas for EuAu$_2$ a precise temperature of 575\,K is required. 
The prepared EuAu$_2$ and GdAu$_2$ ML systems reveal a
$\sqrt{3}\times\sqrt{3}$R30$^{\circ}$ atomic arrangement and a
long range order moir{\'e} lattice with respect to the underlaying Au(111)
substrate. This superstructure is easily distinguishable in low-energy
electron diffraction (LEED) images and serves as a quality indicator in XMCD synchrotron preparations.

The EuAu$_2$ XMCD measurements were performed at Boreas beamline of the
Spanish synchrotron radiation facility ALBA using a 90\% circularly polarized
light from a helical undulator. The measurements were undertaken at 2-20\,K
with a variable magnetic field up to $\pm$6 T. The applied magnetic field $\vec H$
was aligned with the photon propagation vector. The XMCD spectrum is the
difference between the two X-ray absorption spectroscopy (XAS) spectra recorded
with opposite orientation of the magnetic field and/or the circular helicity
of the light, which we call $\mu_+$ and $\mu_-$ for simplicity.
The XMCD signal is proportional to the projection of the magnetization in the
direction of the applied magnetic field. At normal light incidence,
the field is also normal to the sample surface (out-of-plane geometry) while at
grazing incidence (here 70$^{\circ}$), the magnetic field is nearly parallel to the
surface (in-plane geometry). The magnetization curves are taken by varying
continuously the applied field with the sample kept in one of the mentioned
geometries for both circular helicities. For normalization issues, at each field
value two XAS absorption values were taken, the first one corresponding to the photon
energy of the maximum of the XMCD signal, the second one at a slightly lower photon
energy prior to the XAS absorption edge. In order to determine the Curie temperature $T_C$ of the
material such magnetization curves were taken for several temperatures below and above $T_C$.

ARPES experiments were taken at CASSIOPEE beamline of SOLEIL
synchrotron, France, and at our home laboratory in San Sebasti\'an (Spain) together with
the STM and LEED analysis. Photoemission data in San Sebastian were acquired using
Helium I$\alpha$ (h$\nu = 21.2$\,eV) light. In San Sebasti\'an and at SOLEIL a
channelplate-based display type hemispherical analyzer was used (Specs 150 and Scienta R4000
electron analyzers) with angular and energy resolution set to 0.1$^\circ$ and 40\,meV,
respectively. At the synchrotron, $p$-polarized light was used and the sample temperature
during measurements was 70\,K, while sample temperature during Helium I$\alpha$ ARPES
measurements were 120\,K. Resonant photoemission spectroscopy (ResPES) is achieved at the
Eu 4$d\rightarrow$4$f$ absorption edge. In such measurements the photoemission signal
is resonantly enhanced due to a superposition/interference of the direct photoemission
process and an Auger decay and leads to a broad resonant maximum above the 4$d$ absorption
threshold accompanied by a number of narrow peaks caused by several decay processes that was
initially explained for the 4$f^7$ configuration of Gd~\cite{bib:gerken81} and later
for the same configuration of Eu~\cite{bib:sairanen92}. In mixed-valent Eu compounds
slightly different resonant energies~\cite{bib:schneider83} can be used to differentiate
di- and tri-valent contributions.
Here, we used h$\nu = 141$\,eV and 146\,eV corresponding to two close photon energies
in order to minimize photoemission
cross-section changes but still resonantly enhance di- or tri-valent signals.
The off-resonant energy h$\nu = 130$\,eV corresponds to a simple
($4f^7$)$^8S_{7/2}\rightarrow$ ($4f^7$)$^7F_J$ photoemission transition.
The individual $J$ components cannot be resolved easily at such high photon energies due to worse energy
resolution, but have been observed for pure Eu metal at low photon
energies~\cite{bib:kaindl95}.

\subsection{Theoretical Methods}

Density-functional theory (DFT) calculations were performed
in the full-potential linearized augmented plane wave (FLAPW)
formalism \cite{bib:krakauer79,bib:wimmer81,bib:fleur}
at the GGA+U level to describe strong correlation \cite{bib:anisimov97,bib:shick99}
in the full-localized limit approximation,
since the Eu($4f$) orbital is half filled \cite{bib:anisimov93}.
The Perdew-Burke-Ernzerhof (PBE) exchange and correlation functional \cite{bib:pbe96} was used.
The parameter $U=5.5$\,eV is found to match the $4f$ bands binding energies observed in ARPES.
In the case of the calculation parameters for GdAu$_2$, we refer the reader to previous work in
Ref.~\cite{bib:ormaza16}, where $U = 7.5$\,eV was employed.
The lattice constant of the model geometries was fixed to the EuAu$_2$
experimental value. For the supported models a $\sqrt{3}\times\sqrt{3}R30^\circ$
supercell geometry was used with $fcc$ stacking of atomic planes and an interlayer distance of 2.25\,{\AA}
between the flat EuAu$_2$ monolayer and the Au(111) substrate.
For the local FLAPW basis,  as in previous works with REAu$_2$,
RE-$6s,4f,5d$ and Au-$6s,5d$ electrons were included as valence electrons,
and RE-$5s,5p$ and Au-$5p$ as linear orbitals. Partial wave expansions
up to $l_{max}=8$ and 10 were set inside the 
Eu and Au muffin-tin spheres of radii 1.43 and 1.48\,{\AA}, respectively.

SOC was included in the calculation both self-consistently and
in the force theorem perturbative approximation \cite{bib:weinert85,bib:li90,bib:daalderop90,bib:wang93},
i.e. without carrying out further self-consistent optimization of the charge density,
for spins oriented in-plane ($X$) and out-of-plane ($Z$) [see Fig.~\ref{fig:bands}(a)].
Using the force theorem, the MAE is computed as the energy diference between the band energies, i.e. it is
a rigid-band approximation.  The working principles of the method are described in Section~\ref{sec:ft}.
The MAE accuracy is highly dependent on the fine details of the band structure, specially
the gap openings at band crossings and the Fermi level.
Therefore, a fine sampling of
the first Brillouin zone (BZ) and the sharpest possible Fermi-Dirac function are required.
In this work, the MAE was converged with a tolerance of $\sim 0.1$\,meV
using a smearing width $\sigma = 5$\,meV for the Fermi level
(first-order Methfessel-Paxton method \cite{bib:methfessel89}) and a mesh 
of at least $25\times 25$ $k$-points, with 
plane wave expansion cutoffs of 4 and 12 \,bohr$^{-1}$ for
the wavefunctions and potential, respectively.

\section{Results and Discussion}


\begin{figure}[b!]
\centerline{\includegraphics[width=1\columnwidth]{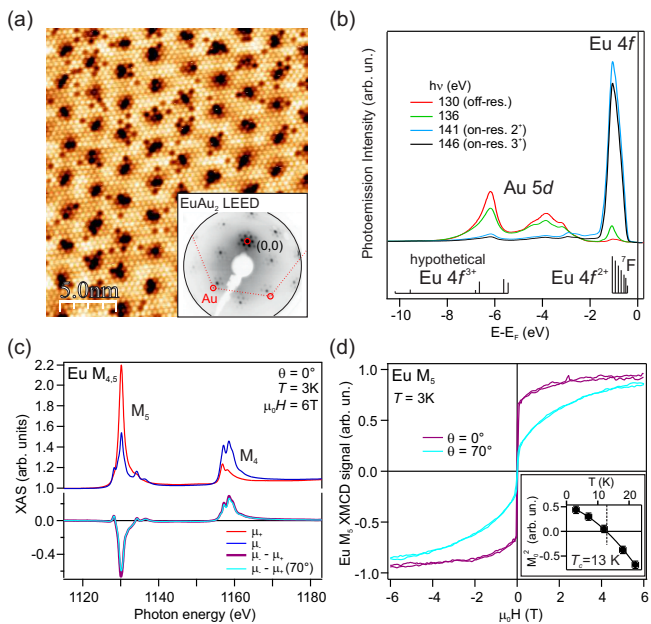}}
\caption{\textbf{Structure, Eu valence and magnetic properties of EuAu$_2$.}
(a) STM image (U = -2V, I = 200 pA). Inset: LEED pattern at $E$ = 40eV
energy, the red circles mark the Au(111) substrate spots prior to EuAu$_2$ formation.
(b) Resonant photoemission across the 4$d\rightarrow$4$f$ absorption
edge determining the pure monolayer formation of divalent Eu atoms. The final state
multiplets are taken from \cite{bib:gerken83,bib:schneider83b}.
(c) X-ray absorption at the Eu M$_{4,5}$ edge with circularly left ($\mu_-$)
and right ($\mu_+$) polarized light at
an applied field of $\mu_0H$ = 6T and at $T$ = 3K in normal incidence
geometry ($\theta$ = 0).
The difference spectrum (XMCD) is shown for
normal and grazing ($\theta = 70^{\circ}$) incidences.
(d) Magnetization curves taken at the Eu M$_5$ edge in these two geometries
and in the inset the corresponding Arrot plot analysis to determine $T_C$.}
\label{fig:stm_res_mag}
\end{figure}

\subsection{Structure Characterization}

The EuAu$_2$ ML is characterized by a hexagonal moir\'e superstructure
visible in the scanning tunneling microscopy (STM) micrograph and 
the LEED pattern of Fig.~\ref{fig:stm_res_mag}(a).
Similar moir\'e superstructures are found
in other RE/NM metal surfaces~\cite{bib:corso10b,bib:ormaza16,bib:fernandez20,bib:que20}.
The STM analysis of EuAu$_2$ yields a moir\'e superlattice constant of
$a_m = (3.3 \pm 0.1)$\,nm with a coincidence lattice of
$(11.5 \pm 0.3)  \times  (11.5 \pm 0.3)$ with respect to the Au(111)
surface ($a_\mathrm{Au} = 0.289$\,nm), in agreement with LEED. The EuAu$_2$ ML
reveals a ($\sqrt{3}\times\sqrt{3}$)R30$^{\circ}$ reconstruction on top of
the Au(111) surface and a lattice parameter of 0.55\,nm.
The $a_m$ of EuAu$_2$ is 10\% smaller than the one found in
GdAu$_2$, HoAu$_2$, or YbAu$_2$,
which show $a_m \approx 3.6$\,nm~\cite{bib:corso10b,bib:fernandez20}.

\subsection{Valency Analysis by ResPES}

The ResPES spectrum 
[see Fig.~\ref{fig:stm_res_mag}(b)]
displays unequivocally the presence
of a single Eu$^{2+}$ multiplet peak and no further peaks related to Eu$^{3+}$
revealing the presence of only divalent Eu atoms
exclusively located at the EuAu$_2$ ML. Eu atoms below the surface (either  di- or trivalent) would give
rise to an additional shifted multiplet~\cite{bib:schneider83} that does not appear in the spectra.
Therefore, the existence of a single EuAu$_2$ ML with divalent character is probed.

\subsection{Magnetic Properties by XAS and XMCD}

Fig.~\ref{fig:stm_res_mag}(c) shows the XAS and XMCD spectra of an EuAu$_2$ ML
at the Eu M$_{4,5}$ absorption edges ($T$ = 3\,K; $\mu_0 H = 6$\,T and normal incidence).
The XAS line shape confirms the divalent character of the Eu atom in EuAu$_2$ ML
deduced from the ResPES spectra of Fig.~\ref{fig:stm_res_mag}(b) [see also Supplemental Material (SM)~\cite{bib:SM}
Fig.~\ref{SM-fig:XAS_Eu_di_trival}
for comparison to XAS spectrum of trivalent Eu$_2$O$_3$].
Magnetization curves of Fig.~\ref{fig:stm_res_mag}(d) are taken at different geometries
and at the photon energy of the maximum M$_5$ XMCD peak while changing $\mu_0 H$.
The out-of-plane magnetization curves with the field applied perpendicular to the surface
saturates at  approx. 2\,T, while the in-plane magnetization curve reaches saturation at
much higher applied fields, close to 6\,T. 
This clearly reveals that the easy axis of magnetization is perpendicular to the plane.
Magnetization loops were recorded at various temperatures,
allowing an Arrot plot estimation~\cite{bib:arrott57,bib:ormaza16} of the Curie temperature
of the surface compound of $T_C = 13$\,K
[see inset of Fig.~\ref{fig:stm_res_mag}(d)].
Further details are found in Fig.~\ref{SM-fig:mag_proc_Eu_Gd}.

The OOP easy axis of magnetization in EuAu$_2$ MLs differs from the
IP one of GdAu$_2$~\cite{bib:fernandez14,bib:ormaza16}, despite of
the identical 4$f$ electronic configuration ($^8S_{7/2}$) and atomic structure.
In the following, we explain this behavior in terms of
their respective band structure measured by ARPES,
which exhibits a clear dependence on the valence character of the REs, namely,
divalent for Eu and trivalent for Gd.
Further analysis by DFT provides  insight in the
spin-orbit effects on the individual bands.

\subsection{Electronic Structure by ARPES and DFT}

\begin{figure*}[tb!]
\centerline{\includegraphics[width=2\columnwidth]{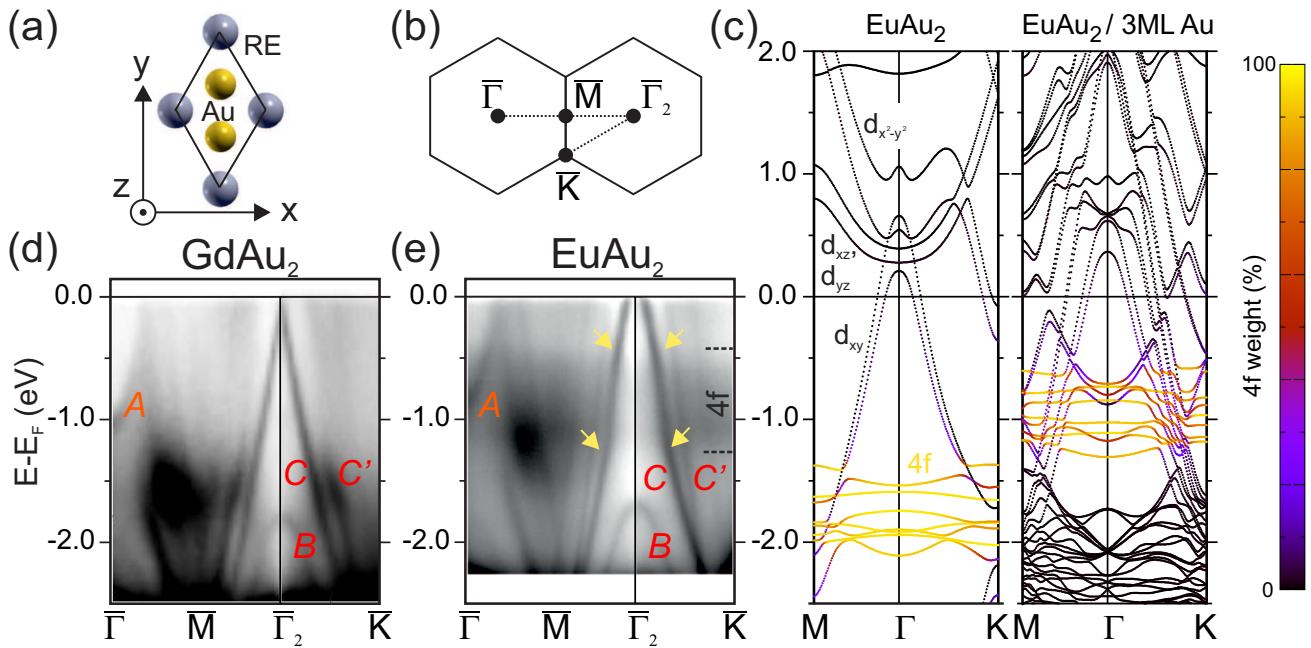}}
\caption{\textbf{Band structures of GdAu$_2$ and EuAu$_2$.}
(a) Model of the REAu$_2$ ML unit cell.
(b) Path in the reciprocal space of the band structures measured by ARPES at photon energy
h$\nu = 21.2$\,eV for GdAu$_2$ (d) and EuAu$_2$ (e).
(c) Band structure from DFT calculations along the indicated high-symmetry direction
for a EuAu$_2$ monolayer, both free-standing (left) and on 3\,ML Au(111) slab-supported (right).
The Eu $4f$ contribution to the states is indicated in a colored scale.}
\label{fig:bands}
\end{figure*}

Results of ARPES measurements performed on EuAu$_2$ MLs show the characteristic
surface dispersive localized bands with Eu($d$)-Au($s,p$) character that
are commonly found in all REAu$_2$ surface compounds
\cite{bib:ormaza13,bib:ormaza16,bib:fernandez20}, see Fig.~ \ref{fig:bands}(d,e) and 
SM Figs.~\ref{SM-fig:ARPESdispersion} and \ref{SM-fig:ARPES}.
Following the notation of Ref.~\cite{bib:fernandez20},
these valence bands are labelled $A$, $B$, $C$, and $C'$.
As already seen in other REAu$_2$ materials,
some of these bands are better detected in the first BZ while
other bands gain intensity in higher order BZ's.
The $4f$ emissions of Eu are seen in the binding energy range between 1.1 and 0.4\,eV,  in
agreement with ResPES of Fig.~\ref{fig:stm_res_mag}(b). 
To better visualize the valence band structure, a low photon energy 
$h\nu = 21.2$\,eV was choosen, where the $4f$ cross-section is relatively low. 
Signs of hybridization between the localized Eu $4f$ level and the EuAu$_2$ valence bands
are observed as kinks in the linear dispersion of the $C$ bands in the
second BZ and are marked with arrows in Fig.~\ref{fig:bands}(e) 
(see also Fig.~\ref{SM-fig:ARPES}).
Such $4f$-valence band hybridizations have been observed for other Eu-based
bulk~\cite{bib:danzenbaecher09} and in YbAu$_2$ surface compounds~\cite{bib:fernandez20}.
The $C$ band in EuAu$_2$ is characterized by a nearly conical shape at $\bar \Gamma$,
with the apex above the Fermi level $E_F$. DFT band structure calculations
for isolated and supported EuAu$_2$ MLs disclose the intersection above
$E_F$ of the $C$ band, of $d_{xy}$ character, with bands of
$d_{x^2-y^2}$ and $d_{xz,yz}$ character, as shown in Fig.~\ref{fig:bands}(c).
These band crossings occur below (above) the Fermi energy for
trivalent (divalent) RE ions~\cite{bib:ormaza16,bib:fernandez20}.
Adding the 3 ML Au(111) substrate slab provide additional dispersive
bands~\cite{bib:ormaza16,bib:fernandez20}, which cause an upward shift of $\approx 0.1$\,eV
in the RE($d$)-Au($s$) hybrid band
manifold, and a strong renormalization of 4$f$ levels. Indeed, in the
calculated free-standing EuAu$_2$ band structure the $4f$ states lie $\approx 0.7$\,eV lower.
The $4f$ band of EuAu$_2$ is half-filled but,
as the individual bands are heavily split by hybridization with the $A$ and $C$ bands, it
shows some degree of dispersion across the BZ. In contrast,
Gd($4f$) orbital in GdAu$_2$ barely interacts with the surrounding metal,
showing a small crystal field splitting and no significant dispersion (see SM Fig.~\ref{SM-fig:euau2_4f_detail}).


\subsection{Force Theorem Analysis}
\label{sec:ft}

SOC at the crossings of the two-dimensional RE($d$)-Au($s$) bands
are sources of anisotropy that we characterize by DFT in the following.
The SOC matrix element for two electrons belonging to orbitals of the same shell is
\begin{equation}
\langle lm \sigma | \xi_l \mathbf{l} \cdot \mathbf{s} | lm' \sigma' \rangle
\label{eq:soc}
\end{equation}
where $\mathbf{l}$ and $\mathbf{s}$ are the one-electron orbital and spin
momentum operators;
$l$, $m$ and $\sigma$ indices label the eigenstates of $\mathbf{l}$, $l_z$, and $\mathbf{s}$, respectively;
and $\xi_l$ is the SOC strength, which is approximately constant for all
the electrons in each $l$-shell. Intershell SOC is negligible.
For hydrogenic atomic orbitals, these matrix elements are analytical
and have been tabulated \cite{bib:abate65,bib:elsasser88}.
Selected combinations of $m$ values and spin orientations give
non-zero matrix elements.
A second-order perturbative treatment of the SOC hamiltonian term shows that
this behaviour is inherited by hybrid bands of an extended system
\cite{bib:takayama76,bib:bruno89,bib:cinal94,bib:vanderlaan98,bib:ke15,bib:sipr16,bib:blanco19}.
In such case, degeneracy lifting at band crossings
obeys the rules imposed by the crystallographic symmetry of the system
and the orbital symmetry of the bands for a given orientation of the spins.

In the so-called force theorem approach, where the SOC correction enters
non-self-consistently in the DFT calculation, the MAE
is obtained from the spin-orientation-dependent band energy contribution
to the total energy of the system \cite{bib:weinert85,bib:daalderop90}:
\begin{equation}
\mathrm{MAE} = \sum_{kn} \epsilon_{kn}^{X} f_{FD}( \epsilon_{kn}^{X}-E_F^{X}) -
 \sum_{kn} \epsilon_{kn}^{Z}  f_{FD}( \epsilon_{kn}^{Z} - E_F^{Z} )
\label{eq:maeft}
\end{equation}
where the sum runs over the eigenenergies $\epsilon_{kn}^{\hat s_a}$ 
($k$ and $n$ are the $k$-point and band indices, respectively), 
$f_{FD}$ is the Fermi-Dirac function, and $E_F^{\hat s_a}$ are the Fermi energies
for the spin orientations $\hat s_a=Z$ (out-of-plane) and $X$ 
[the in-plane direction along nearest RE atoms, shown in Fig.~\ref{fig:bands}(a)]. 
Azimuthal dependence of the MAE is small.
According to this equation, non-zero contributions to the MAE are
generated at band crossings that become gapped for one
spin orientation, but not for the other.
The final easy-axis direction results from the integration of all individual gap
contributions, of positive and negative small values, over occupied states.
At crossings of fully occupied bands,
the energy dispersion around the degeneracy point has to
be asymmetrical in order to have a sizable contribution to MAE,
otherwise band energies around the gapped feature will
cancel out each other. If the degeneracy lifting occurs
exactly at the Fermi level, the contribution to the MAE will be 
large \cite{bib:daalderop90,bib:daalderop94}.

A spectral analysis of Eq.~\ref{eq:maeft} 
is not possible, as the Fermi energy depends on the spin orientation.
Instead, the common practice is to plot MAE {\it vs.} $N_e$, i.e.
$\mathrm{MAE}(N_e)$, where $N_e$ is the number of electrons
with respect to the neutral case ($N_e=0$) calculated as an integral
over the density of states of the system for each spin orientation, i.e.
each $N_e$ value provides $E_F^{X,Z}(N_e)$ values
\cite{bib:daalderop94,bib:moos96,bib:lessard97,bib:ravindran01,bib:blanco19}.
Therefore, the $\mathrm{MAE}(N_e)$ curve allows to asign a positive or a negative MAE contribution to
the region of the energy range close to $E_F^{X,Z}(N_e)$.
As the Fermi energy fluctuation with the spin direction is small (a few meV),  
we use the $E_F(N_e)$ value calculated without SOC to guide the eye in the graphs.
The value obtained at neutrality, MAE($N_e=0$), corresponds to the expected anisotropy of the system.
This type of MAE analysis relies on the validity of the force theorem approximation,
namely, it assumes that there are only minor
changes in the electron density (essentially, loss of collinearity) 
caused by the effect of SOC \cite{bib:li90,bib:daalderop90}.
Despite the two REAu$_2$ compounds under study being heavy-atom systems,
by comparing to fully self-consistent calculations of the band structures with SOC,
we find that the force theorem properly accounts for the
spin-orbit induced gaps in an energy range of several eV around the Fermi
energy (see SM Figs.~\ref{SM-fig:gdau2_scf_vs_ft} and \ref{SM-fig:euau2_scf_vs_ft}
for GdAu$_2$ and EuAu$_2$, respectively).

\subsubsection{GdAu$_2$}

\begin{figure*}[tb!]
\centerline{\includegraphics[width=2\columnwidth]{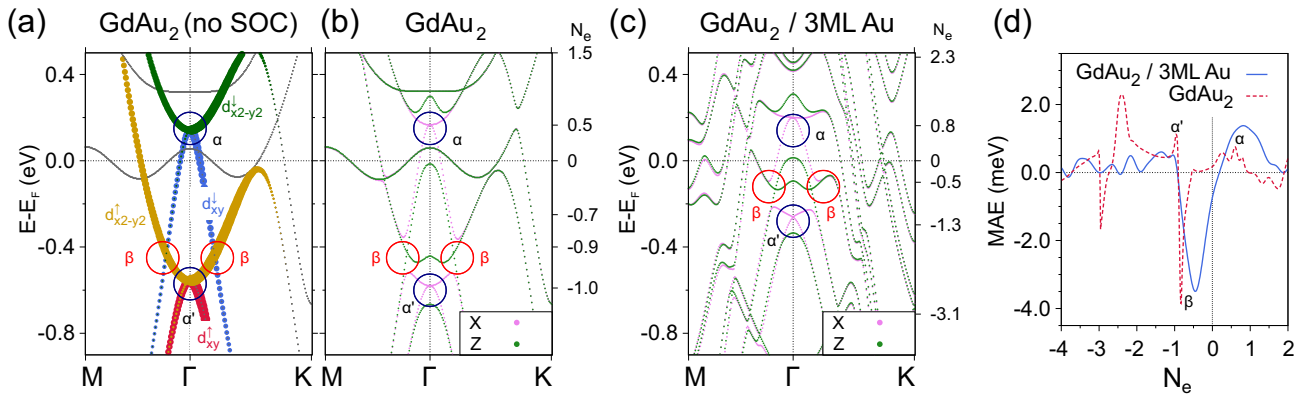}}
\caption{\textbf{Band structure and MAE of GdAu$_2$ as a function of band filling.}
(a) Orbital and spin characters of free-standing GdAu$_2$ bands
in the absence of SOC.
Band structure of a free-standing (b) and supported (c) GdAu$_2$ ML for spins aligned
to the X (violet) and Z (green) directions in the force theorem approximation,
each calculation referred to its own Fermi level.
The SOC-induced gaps at the band crossings relevant to the MAE
are indicated with circles labelled $\alpha,\alpha'$ and $\beta$.
The right-hand side axes show the corresponding band filling,
i.e. the number of electrons $N_e$ referred to charge neutrality.
(d) MAE as a function of $N_e$ for free-standing (solid)
and supported (dashed) GdAu$_2$.
Positive (negative) MAE values mean OOP (IP) easy magnetization axis.
}
\label{fig:kelly_gdau2}
\end{figure*}

We apply this analysis to the band structure of free-standing and supported GdAu$_2$ MLs,
shown in Fig.~\ref{fig:kelly_gdau2}(b-c).
The band crossings labelled $\alpha$ and $\alpha'$ in the figures
are degenerate for in-plane magnetization, but
split by OOP magnetization.
Fig.~\ref{fig:kelly_gdau2}(a) panels shows, as a guide for the eye, the orbital
characters and spin polarities in the absence of SOC for free-standing GdAu$_2$.
The $\alpha$ and $\alpha'$ features are crossings between $d_{xy}$ and $d_{x^2-y^2}$
($m=\pm 2$) with \emph{equal} spin polarization, a situation where the matrix element Eq.~\ref{eq:soc}
foresees splitting by OOP magnetization.
The Weyl nodal line $\beta$ \cite{bib:feng19},
which is a ring-shaped crossing around the $\Gamma$ point
between  $d_{xy}$ and $d_{x^2-y^2}$ bands
with \emph{opposite} spin polarization [see Fig.~\ref{fig:kelly_gdau2}(a)],
is split by in-plane magnetization according to Eq.~\ref{eq:soc}.

Fig.~\ref{fig:kelly_gdau2}(d) shows the obtained $\mathrm{MAE}(N_e)$.
The curves $E_F(N_e)-E_F(0)$ without SOC, shown in the SM Fig.~\ref{SM-fig:filling_reference}
establish an approximate mapping between band fillings and binding energies 
[a few discrete values are given also in the right-hand side axes of Fig.~\ref{fig:kelly_gdau2}(b,c)].
For GdAu$_2$ close to neutrality ($N_e=0$) the
essential contributions to the MAE are those of $\alpha,\alpha'$ and $\beta$ features.
In the free-standing case, the $\beta$ nodal ring results in the net sharp
MAE peak of negative values in the dashed red line at $N_e=-0.9$ (i.e. contributing to IP 
easy axis of magnetization),
whereas features $\alpha$ and $\alpha'$ contribute to the
OOP easy magnetization axis with peaks of positive MAE at $N_e=-1$ and just above $N_e=0$
in the red dashed curve of Fig.~\ref{fig:kelly_gdau2}(d).
At neutrality we obtain an OOP easy axis with $\mathrm{MAE}(0)=0.17$\,meV.

The same features $\alpha,\alpha'$ and $\beta$ appear for GdAu$_2$/3\,ML Au(111),
albeit at binding energies higher by $\sim 0.1$\,eV [see Figs.~\ref{fig:kelly_gdau2}(b,c)],
which also result in positive and negative contributions to the MAE around the
neutrality point. The $\beta$ line causes the in-plane peak at $N_e=-0.5$ (solid blue line),
and $\alpha$ and $\alpha'$ contribute to the OOP magnetization easy axis with positive-valued 
MAE peaks  at $N_e=-1.3$ and 0.8, respectively.
At the neutrality point we find $\mathrm{MAE}(0)=-0.9$\,meV,
i.e. IP easy axis of magnetization, same as observed experimentally.
This is due to the contribution of the $\beta$ gapped line, which
dominates over $\alpha$ and $\alpha'$ single-point gaps at $\Gamma$, aided by the
fact that the $\beta$ gapped line lies closer to $E_F$ in the supported case.
The contribution of the GdAu$_2$ $4f^7$ electrons to the anisotropy lies within the force theorem error,
since the departure from sphericity of this orbital is negligible when embedded in the alloy
(in other words, the Gd($4f$) spin density cloud would be almost insensitive to
the applied field orientation, as shown in SM Fig.~\ref{SM-fig:euau2_4f_detail}).
Therefore, we attribute the experimental easy magnetization plane essentially
to the MAE contribution of the dispersive Gd($d$)-Au($s$) bands.

\subsubsection{EuAu$_2$}

\begin{figure}[tb!]
\centerline{\includegraphics[width=1\columnwidth]{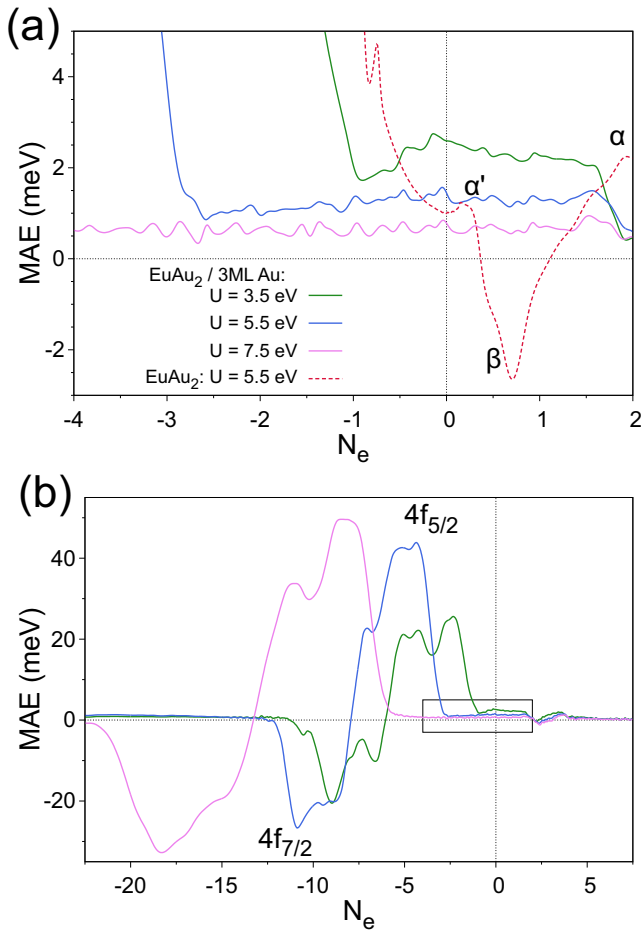}}
\caption{\textbf{MAE of EuAu$_2$ as a function of band filling}
(a) MAE as a function of $N_e$ for free-standing (solid blue)
and supported (dashed red) EuAu$_2$ calculated with $U=5.5$\,eV
close to the neutrality condition $N_e=0$.
Positive (negative) MAE values contribute to the OOP (IP) easy
magnetization axis.
Additional solid curves show the results for EuAu$_2$/3\,ML Au(111)
with other $U$ values.
(b) Same curves in a wider $N_e$ to show the large contributions
of $4f$ electrons, switching the easy axis of magnetization
from IP (negative MAE) to OOP (positive MAE).
The rectangle corresponds to the close-up shown in panel (a).}
\label{fig:kelly_euau2}
\end{figure}

Next, we apply the force theorem methodology to the EuAu$_2$ bands.
Bare inspection of the band structures of both
free-standing and supported EuAu$_2$ MLs rules out the possibility
of the anisotropy being dominated by the SOC effects on the Eu($d$)-Au($s$) hybrid
bands for two reasons:
(i) due to the divalent character of Eu,
the  $d_{xy}$ and $d_{x^2-y^2}$ band crossings relevant for MAE
are unoccupied [see Fig.~\ref{fig:bands}(c)]
and thus they cannot contribute to the observed anisotropy, and
(ii) the $f$-band is heavily broadened by hybridization, with
individual bands showing significant dispersion and spin orientation dependence
[see SM Fig.~\ref{SM-fig:euau2_4f_detail}(d)].

The force theorem Eq.~\ref{eq:maeft} applied to the $U=5.5$\,eV band structures yields
an OOP easy magnetization axis both for free-standing and supported EuAu$_2$, with
$\mathrm{MAE}(0)=1.0$ and 1.54\,meV, respectively
[these are, respectively, the values taken by the thick red dashed and blue solid curves, 
at neutrality in Fig.~\ref{fig:kelly_euau2}(a)].
The MAE peaks due to $\alpha',\beta$ and $\alpha$ band crossing features
are visible in the free-standing EuAu$_2$ bands
above the charge neutrality level (see also Fig.~\ref{SM-fig:euau2_bands}).
For EuAu$_2$/3\,ML Au(111), Fig.~\ref{fig:kelly_euau2}(b) shows two large broad peaks
at each $\mathrm{MAE}(N_e)$ curve,
which take values $\simeq \pm 0.1$\,eV at the electron filling ranges
that correspond to the binding energy ranges where the hybridized $4f$ bands lie.
The peak pairs account for the filling of the
SOC-split band manifolds $4f_{7/2}$ and $4f_{5/2}$.
In the calculation with $U=5.5$\,eV,
these peaks occur at $N_e=-10$ and $-5$, respectively, corresponding
to binding energies close to -1\,eV
[see also SM Fig.~\ref{SM-fig:filling_reference}(b,c)].
These bands yield a non-negligible positive residual
contribution to $\mathrm{MAE}(0)$ by virtue of the $f-d$ hybridization.
In a single-ion picture, this would be interpreted as a loss of sphericity of the
half-filled $4f$ shell, which acquires a net non-zero $\langle L \rangle$ \cite{bib:rinehart11}.
The SM Fig.~\ref{SM-fig:euau2_4f_detail} shows the magnetization
distribution projected on $\hat s_a$ [$m_{\hat s_a}(\mathbf{r})$]
of the Eu($4f$) shell embedded in the free-standing alloy monolayer,
obtained by integration of the Kohn-Sham states in the binding energy range between -2.25 and -1.25\,eV.
The magnetization anisotropy distribution $m_X(\mathbf{r}) - m_Z(\mathbf{r})$
takes positive and negative values in-plane and out-of-plane, respectively,
which can be interpreted as the Eu($f$) orbital spin density tendency to become deformed
along the applied field direction.
Note that the same quantity integrated for Gd($f$) between -10 and -8.5\,eV is
an order of magnitude smaller.

The $4f$ contribution to the MAE of EuAu$_2$
can be further probed by modifying the $U$ correlation parameter.
We have carried out force theorem analyses for supported EuAu$_2$ with $U=3.5$ and $7.5$\,eV.
When $U$ is decreased, the MAE energy is increased by $\simeq 1$\,meV and
when $U$ is increased, the easy axis of magnetization remains out-of-plane, 
but the $\mathrm{MAE}(0)$ value is reduced (see Fig.~\ref{fig:kelly_euau2}).
The densities of states [SM Fig.~\ref{SM-fig:filling_reference}(d)] show that
a larger $U$ value implies less hybridization between the $4f$ electrons and
the hybrid bands of $d$ character crossing the Fermi level.
Indeed, for $U=7.5$\,eV the $4f$ band lies below $E_F-1.5$\,eV, i.e. close to the
Au substrate states and below the $d_{xy}$ band. For $U=3.5$\,eV, however,
the $4f$ band is centred at $E_F-0.5$\,eV, producing hybrid $f-d$ states close to the Fermi level.
This shows that the degree of $f-d$ hybridization drives the MAE behaviour of EuAu$_2$.

\section{Conclusions}

To summarize, our XMCD experiments have determined that
a EuAu$_2$ ML on Au(111) shows out-of-plane easy magnetization axis.
Using photoemission experiments combined with DFT calculations,
we show that this is a consequence of the Eu$^{2+}$ valence state and the $f-d$ band hybridization.
In contrast, the easy magnetization plane of GdAu$_2$ ML is explained by Gd$^{3+}$ valence and spin-orbit
splitting of the point and line degeneracies of bands with $d_{xy,x^2-y^2}$ character.
The role of the RE valence is to determine whether these specific states are occupied or empty.
This model of the contribution of the itinerant electrons to the magnetocrystalline
anisotropy is a general result for the REAu$_2$ family of monolayer compounds.
For RE=Eu,Gd this is the only mechanisms at work, since the $4f$ orbital is
half-filled, i.e. $L=0$. 
For RE atoms with non-zero $L$ quantum numbers, there is a single-ion anisotropy 
that results from the interplay between spin-orbit and crystal-field splittings
of the $4f$ multiplet. For example, this multiplet mechanism dominates the OOP easy 
magnetization axis behaviour in HoAu$_2$, where the valence is Ho$^{3+}$ \cite{bib:fernandez20}. 
We predict the itinerant electrons to yield MAE values of $\approx 1$\,meV in REAu$_2$ 
monolayers, irrespective of the $L$ value. 
Therefore, the band and multiplet mechanisms may eventually compete.

\begin{acknowledgments}
Discussions with the late J.I. Cerd\'a are warmly thanked.
Financial support from projects MAT-2017-88374-P, PID2020-116093RB-C44 and
PID2019-103910GB-I00, funded by MCIN/AEI/10.13039/501100011033/,
the Basque Government (grants IT-1255-19, IT1260-19), and
the University of the Basque Country (UPV/EHU) (grant GIU18/138) is acknowledged.
Computational resources were provided by DIPC.
The SOLEIL based research leading to the results has been supported 
by the project CALIPSOplus under Grant Agreement 730872 
from the EU Framework Programme for Research and Innovation HORIZON 2020.
L.F. acknowledges financial support from the
EU's Horizon 2020 research and innovation programme under the
Marie Sk\l odowska-Curie grant agreement MagicFACE No 797109.
\end{acknowledgments}

\clearpage



\bibliography{maere}

\makeatletter\@input{aux4_EuAu2_GdAu2_SI_250122.tex}\makeatother

\end{document}


\newcommand{\ALBA}[0]{{
ALBA Synchrotron Light Source, Carretera BP 1413 km 3.3, 08290 Cerdanyola del Vall{\`e}s, Spain}}
\newcommand{\CFM}[0]{{
Centro de F\'{\i}sica de Materiales CSIC/UPV-EHU-Materials Physics Center, E-20018 San Sebasti\'an, Spain}}
\newcommand{\UPVPMA}[0]{{
Departamento de Pol\'imeros y Materiales Avanzados: F\'isica, Qu\'imica y Tecnolog\'ia,
Universidad del Pa\'is Vasco UPV-EHU, 20018 San Sebasti\'an, Spain}}
\newcommand{\DIPC}[0]{{
Donostia International Physics Center, E-20018 Donostia-San Sebasti\'an, Spain}}
\newcommand{\UPVFA}[0]{{
Departamento de F\'isica Aplicada I,
Universidad del Pa\'is Vasco UPV/EHU,  E-20018 San Sebasti\'an, Spain}}
\newcommand{\IOMCNR}[0]{{
IOM-CNR, Laboratorio TASC, Strada Statale 14 Km 163.5, I-34149 Trieste, Italy}}
\newcommand{\Elettra}[0]{{
Elettra - Sincrotrone Trieste S.C.p.A., Strada Statale 14 Km 163.5, I-34149 Trieste, Italy}}
\newcommand{\TUDD}[0]{{
Institut f\"ur Festk\"orper- und Materialphysik, Technische Universit\"at
Dresden, D-01062 Dresden, Germany}}
\newcommand{\Soleil}[0]{{
Synchrotron SOLEIL, CNRS-CEA, L'Orme des Merisiers, Saint-Aubin-BP48, 91192 Gif-sur-Yvette, France}}
\newcommand{\ISMTrieste}[0]{{
Instituto di Struttura della Materia, Consiglio Nazionale delle Ricerche, 34149 Trieste, Italy}}

\author{M. Blanco-Rey}
\affiliation{\UPVPMA}
\affiliation{\DIPC}
\author{R. Castrillo-Bodero}
\affiliation{\CFM}
\author{K. Ali}
\affiliation{\DIPC}
\affiliation{\CFM}
\author{P. Gargiani}
\affiliation{\ALBA}
\author{F. Bertran}
\affiliation{\Soleil}
\author{P. M. Sheverdyaeva}
\affiliation{\ISMTrieste}
\author{J.E. Ortega}
\affiliation{\DIPC}
\affiliation{\CFM}
\affiliation{\UPVFA}
\author{L. Fernandez}
\affiliation{\CFM}
\affiliation{\UPVFA}
\author{F. Schiller}
\affiliation{\DIPC}
\affiliation{\CFM}
\email{frederikmichael.schiller@ehu.es}

\title{Supplemental Material for:
``Valence state determines the band magnetocrystalline anisotropy in 2D rare-earth/noble-metal compounds''}

\date{\today}

\maketitle


\section{Additional photoemission results}


\begin{figure}[h]
\centerline{\includegraphics[width=0.7\columnwidth]{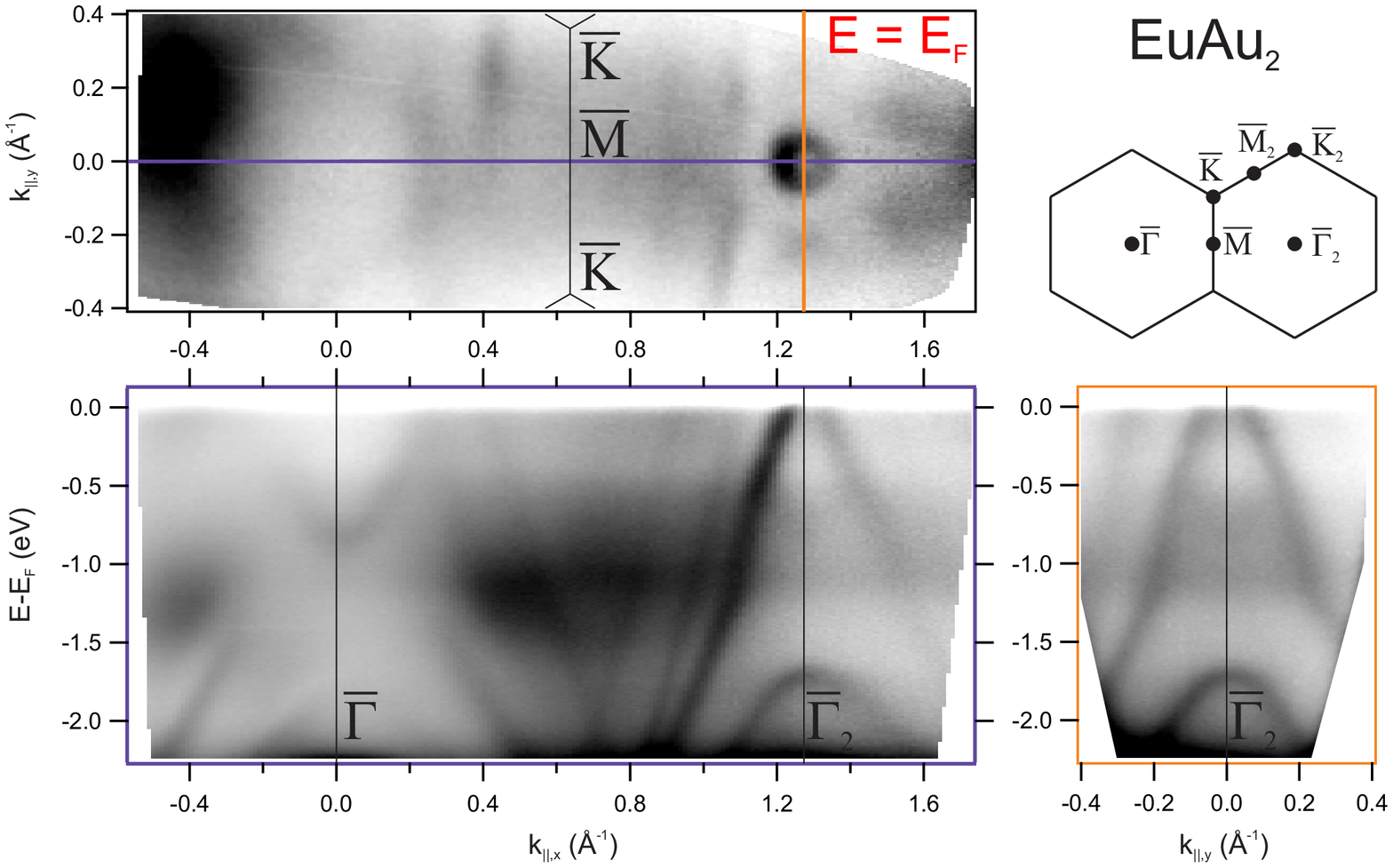}}
\caption{\textbf{ARPES intensity mappings of EuAu$_2$ at h$\nu$ = 21.2\,eV. }
Fermi energy intensity plot (Fermi surface mapping) of the EuAu$_2$
surface and band structure cuts along the $\bar\Gamma\bar M\bar\Gamma_2$ and
$\bar K_2\bar\Gamma_2\bar K_2$ surface Brillouin zone directions. }
\label{fig:ARPESdispersion}
\end{figure}

\begin{figure}[h]
\centerline{\includegraphics[width=0.28\columnwidth]{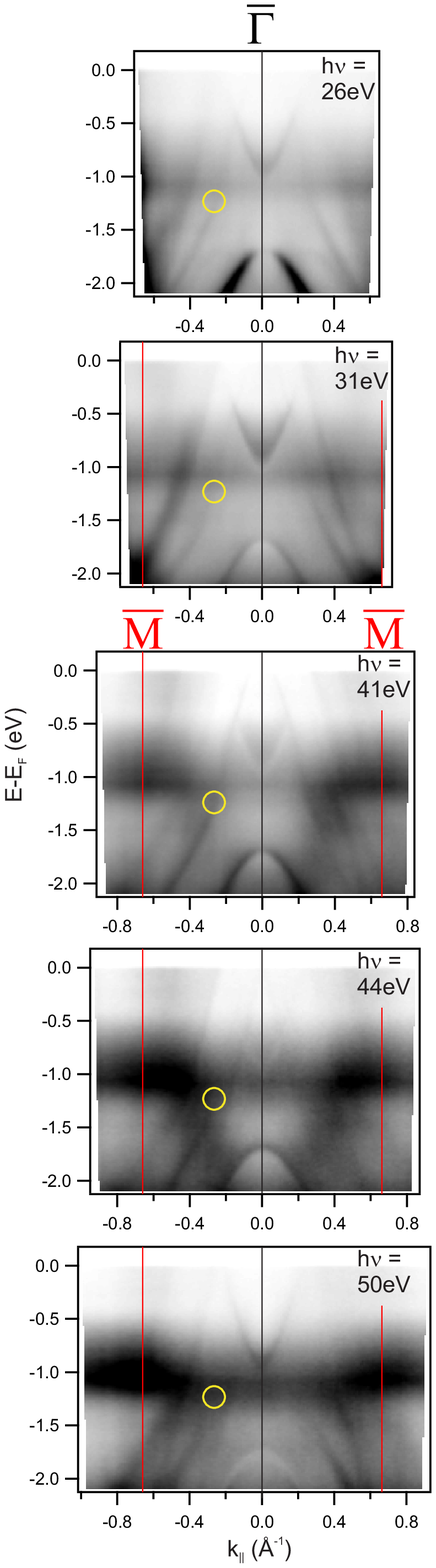}}
\caption{\textbf{ARPES intensity plots along the $\bar M\bar\Gamma\bar M$ high symmetry direction in EuAu$_2$.}
Photoemission intensity mappings along the $\bar\Gamma_2\bar M\bar\Gamma\bar M\bar\Gamma_2$ high symmetry line.
The yellow circle at -0.27\,{\AA} and -1.2\,eV reveal that the observed kink in the C band does not disperse for 
different photon energies.}
\label{fig:ARPES}
\end{figure}

\begin{figure}[h]
\centerline{\includegraphics[width=0.9\columnwidth]{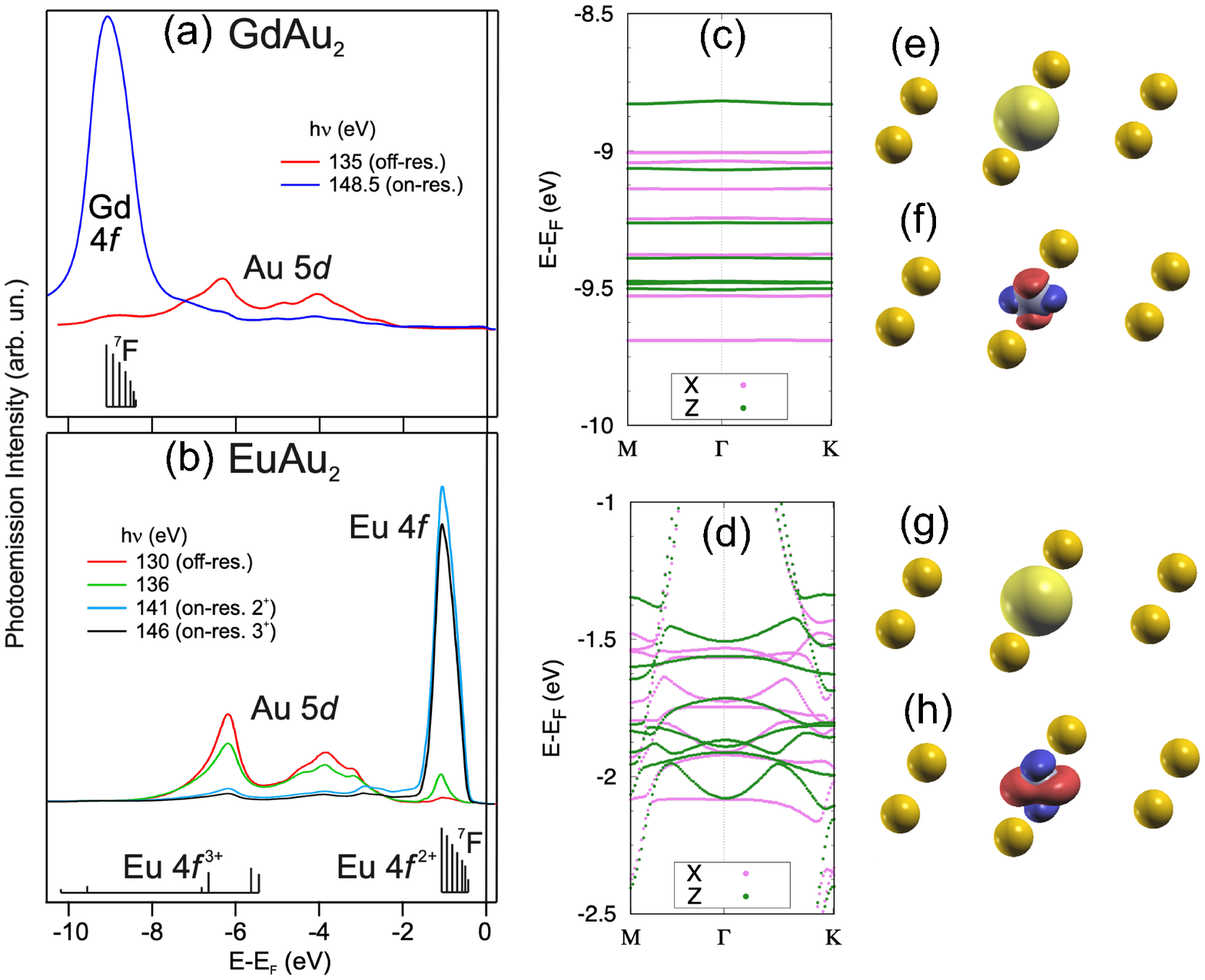}}
\caption{\textbf{Electronic structure of the $4f$ states in free-standing GdAu$_2$ and EuAu$_2$.}
(a,b) Off- and on-resonant photoemission of GdAu$_2$ and EuAu$_2$, 
respectively, obtained at the indicated photon energies. 
The indicated multiplets are taken from Refs.~\cite{bib:gerken83,bib:schneider83}. 
(c,d) Detail of the self-consistently calculated spin-orbit-split $4f$ bands of free-standing
GdAu$_2$ (c) and EuAu$_2$ (d) for spins oriented along
the $X$ (purple) or $Z$ (green) directions.
Note the contrast between the weak dispersion and strong hybridization
of the GdAu$_2$ and EuAu$_2$ cases, respectively.
(e,g) Real-space representations of the RE($4f$) charge density $\rho (\mathbf{r})$,
obtained from the integration of the Kohn-Sham states with
spins along the $Z$ direction shown in panels (c) and (d),
respectively. The depicted isoelectronic surface is 0.1\,$e a_0^{-3}$.
In both atoms, the deviation from sphericity are negligible by bare eye.
The next quantity can, in contrast, probe the asphericity of the orbital with more precision.
(f,h) Real-space representations of the anisotropy of the $4f$
magnetization density, calculated as $\Delta m  (\mathbf{r}) =  m_X(\mathbf{r})-m_Z(\mathbf{r}) $,
with red (blue) indicating $\Delta m  (\mathbf{r}) > 0$ ($<0$).
The depicted isosurfaces in this case are 0.0001\,$e a_0^{-3}$ for Gd($4f$) (g) and
0.001\,$e a_0^{-3}$ for Eu($4f$) (h).
The $m^\mathrm{Gd}(\mathbf{r})$  sensitivity  to the magnetization axis is
smaller than that of  $m^\mathrm{Eu}(\mathbf{r})$ by an order of magnitude,
which means that Eu($4f$) is more prone to be deformed by a magnetic field, i.e.
it can provide non-negligible contributions to the MAE by departure of its spherical shape,
as shown in the main paper Fig.~\ref{MP-fig:kelly_euau2}.
}
\label{fig:euau2_4f_detail}
\end{figure}

\newpage
\clearpage

\section{Additional XAS results}

\begin{figure}[h]
\centerline{\includegraphics[width=0.7\columnwidth]{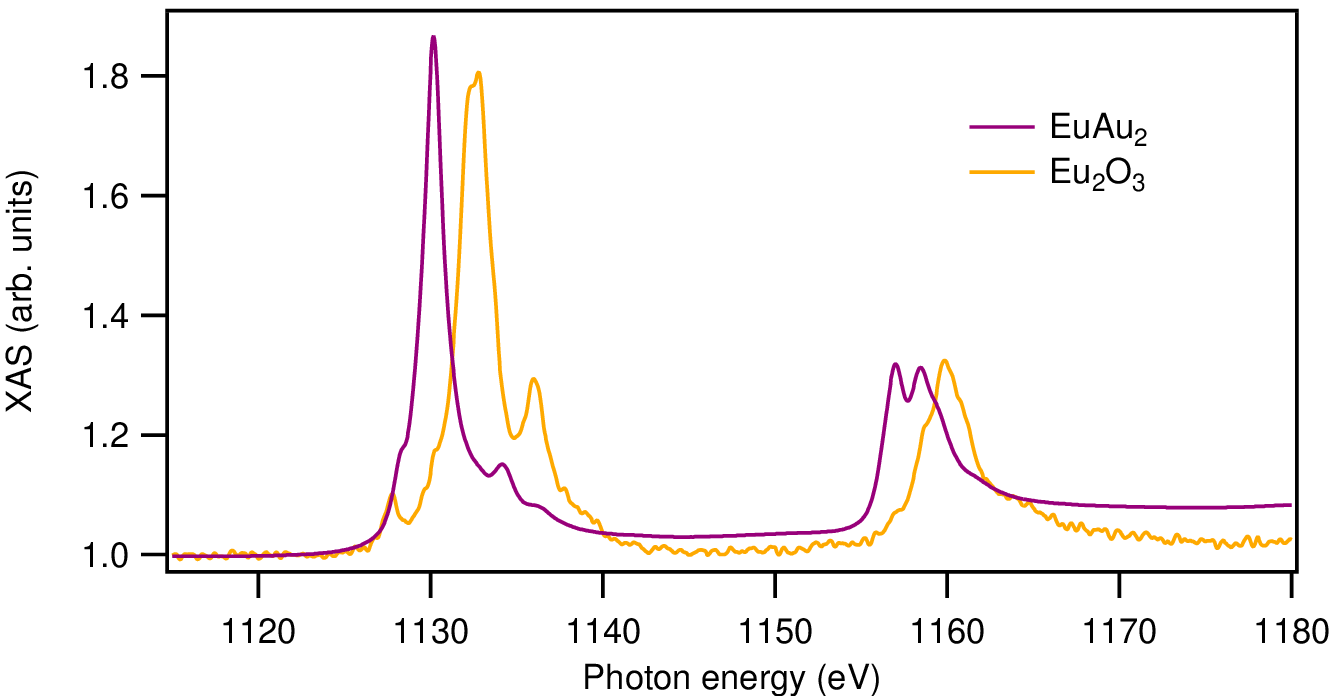}}
\caption{\textbf{X-ray absorption spectra taken with linear horizontal polarized light
at normal incidence geometry for divalent EuAu$_2$ and trivalent Eu$_2$O$_3$.}
X-ray absorption spectroscopy results can be used to easily distinguish
di- and tri-valent Eu atoms in compounds. Here we
show the spectra of completely divalent Eu in EuAu$_2$ and trivalent Eu in Eu$_2$O$_3$.
For comparison, theoretical and experimental spectra of divalent Eu atoms in different 
compounds can be found in Refs.~\cite{bib:thole85,bib:kinoshita02,bib:tcakaev20}.}
\label{fig:XAS_Eu_di_trival}
\end{figure}

\begin{figure}[h]
\centerline{\includegraphics[width=0.9\columnwidth]{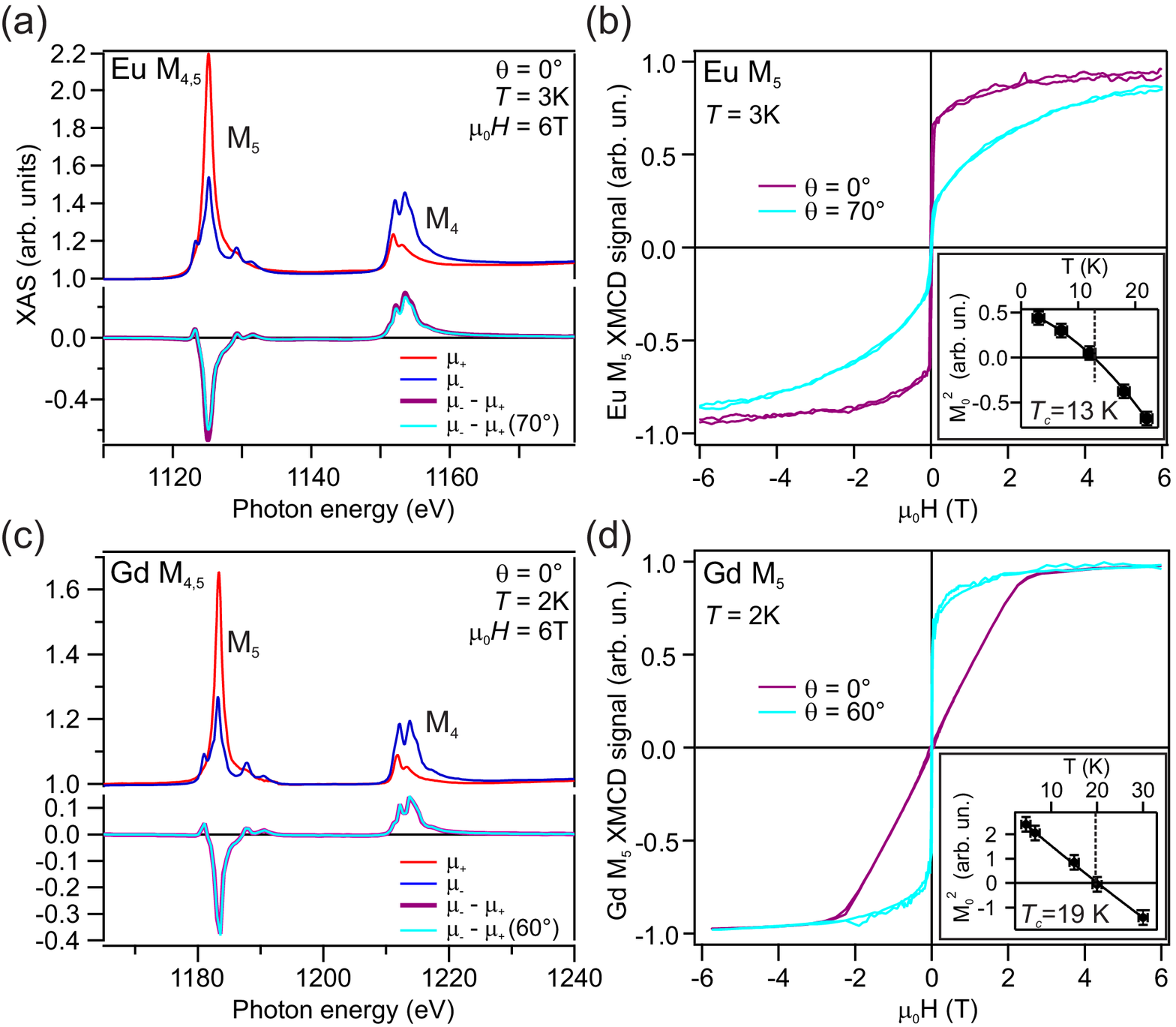}}
\caption{\textbf{Comparison of the magnetic properties of EuAu$_2$ and GdAu$_2$.}
(a),(c) X-ray absorption spectra taken with left and right polarized light at $T$ = 3K
and with an applied field of $\mu_0H$ = 6T with the field applied normal to the surface
for EuAu$_2$ and GdAu$_2$, respectively. The resulting difference, the XMCD signals,
are shown in the bottom part for fields normal to the surface $\theta$ = 0$^{\circ}$,
at $\theta$ = 70$^{\circ}$ and 60$^{\circ}$, respectively. (b),(d) The
magnetization curves are obtained by acquisition of the XAS signal at the maximum
of the XMCD signal (and a pre-edge value for normalization) varying continually the
applied field from +6T to -6T and then from -6T to +6T for both circular light polarizations.
The Curie temperature $T_C$ can be extracted by the Arrott plot
analysis~\cite{bib:arrott57,bib:ormaza16} of magnetization curves taken
at several temperatures around $T_C$. As a result we obtain $T_C$ = 13K and 19K for
the EuAu$_2$ and GdAu$_2$ surface compounds, respectively. Most important here is
the change in the anisotropy, resulting in an out-of-plane easy axis of magnetuzation for Eu atoms
in EuAu$_2$ and an in-plane easy axis for Gd in GdAu$_2$.}
\label{fig:mag_proc_Eu_Gd}
\end{figure}

\clearpage

\section{Tests on the validity of the force theorem approach}

\begin{figure}[h]
\centerline{\includegraphics[width=0.7\columnwidth]{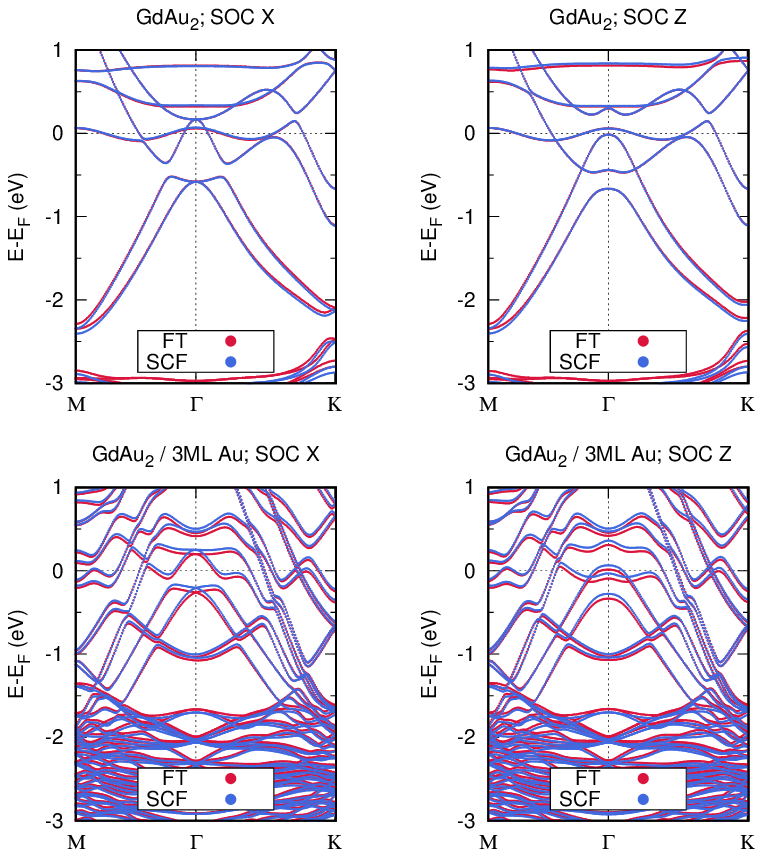}}
\caption{\textbf{Self-consistent and force-theorem bands for free-standing and supported GdAu$_2$ structures.}
Band structures of free-standing and supported GdAu$_2$ are shwon for spins aligned
to in-plane (panels labelled as SOC X) and out-of-plane (SOC Z) directions.
The eigenenergies obtained for SOC included in the force theorem approximation
(labelled FT, red) are compared with the self-consistent ones (SCF, blue).
Each individual calculation is referred to its own Fermi level
(note that the Fermi energy changes if the spin orientation changes).
Agreement between FT and SCF calculations in the case of a 3\,ML Au substrate is
reduced with respect to that of free-standing GdAu$_2$
due to the intense SOC at the substrate bands, which
introduces small overall shifts in the band structure. Despite this,
the overall band structure, and in particular the gap openings,
for the two spin directions are satisfactorily reproduced by the force theorem method.}
\label{fig:gdau2_scf_vs_ft}
\end{figure}

\begin{figure}[h]
\centerline{\includegraphics[width=0.7\columnwidth]{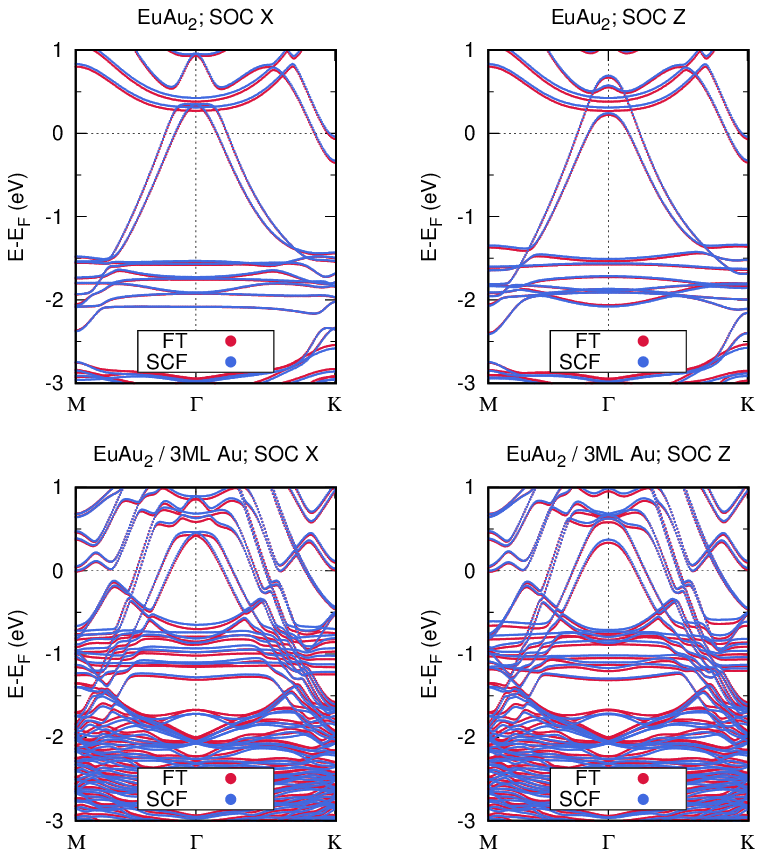}}
\caption{\textbf{Self-consistent and force-theorem bands for free-standing and supported EuAu$_2$ structures.}
Same information as in Fig.~\ref{fig:gdau2_scf_vs_ft} for the EuAu$_2$ case.
}
\label{fig:euau2_scf_vs_ft}
\end{figure}

\clearpage

\begin{figure}[h]
\centerline{\includegraphics[width=0.7\columnwidth]{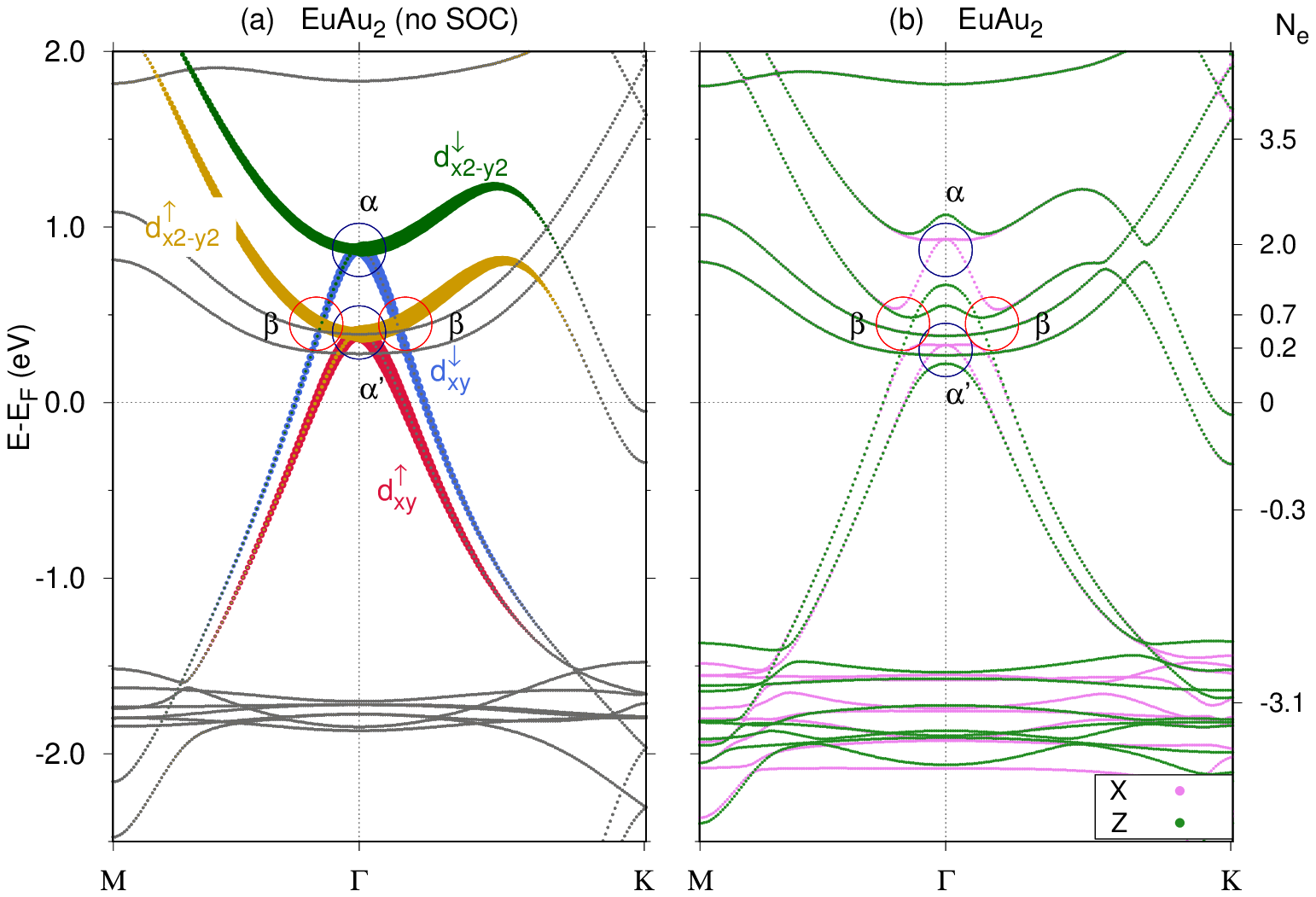}}
\caption{\textbf{Band structure for free-standing EuAu$_2$.}
(a) To guide the eye, orbital and spin characters of free-standing EuAu$_2$ bands
in the absence of SOC.
(b) Band structure of free-standing EuAu$_2$ with SOC (force theorem) for 
spins aligned to the X (violet) and Z (green) directions.  
The right-hand side axis shows the corresponding band filling $N_e$ 
(see also Fig.~\ref{fig:filling_reference}). Note that the gap openings 
that were relevant in the anisotropy of trivalent GdAu$_2$ (labelled $\alpha, \alpha'$ and $\beta$) 
lie above the Fermi level in the EuAu$_2$ case and therefore do not contribute to the observed MAE.
}
\label{fig:euau2_bands}
\end{figure}

\clearpage

\section{Visual guide to interpret the MAE($N_e$) curves}

\begin{figure}[h]
\centerline{\includegraphics[width=0.9\columnwidth]{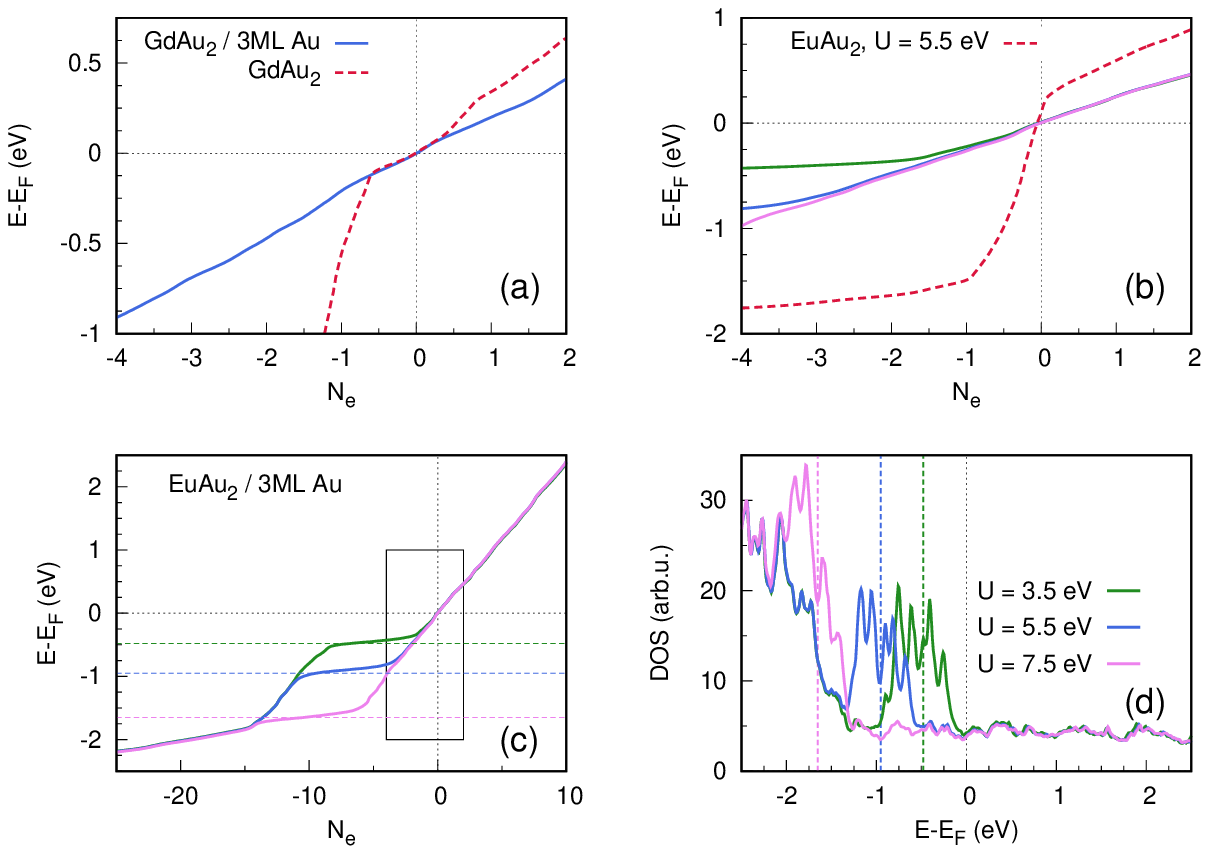}}
\caption{\textbf{Correspondence between $N_e$ and Fermi energies.
Dependence of EuAu$_2$ / 3\,ML Au electronic structure and MAE with the correlation parameter.}
(a-c) Guides to the eye to interpret the MAE($N_e$) curves in the main paper.
The curves represent the Fermi energy that corresponds to a
given filling with respect to neutrality ($N_e=0$),
as in the right-hand axis tics of main paper Figs.~\ref{MP-fig:kelly_gdau2}(b,c).
These fillings are calculated as an integral of the
density of states for each one of the four considered systems without including SOC.
(a) free-standing and supported GdAu$_2$ to interpret the MAE($N_e$)
curves of Fig.~\ref{MP-fig:kelly_gdau2}(d).
(b) Same for EuAu$_2$ [Fig.~\ref{MP-fig:kelly_euau2}(a)].
(c) A wider range of fillings allows to identify the Eu($4f$) contributions
in Fig.~\ref{MP-fig:kelly_euau2}(b), located at different binding energies
(indicated by horizontal dashed lines) or $N_e$ interval (plateaus)
depending on the $U$ values.
The $4f$ broadened band is displaced toward the Fermi level for lower $U$ values.
(d) Total density of states calculated for EuAu$_2$ / 3\,ML Au with spins perpendicular to the surface
in the force theorem approximation for three values of the correlation parameter $U$.
For $U=5.5$\,eV (blue), the broad peak centered at energies $E-E_F \approx -1$\,eV corresponds
to the hybridized $4f$ bands, in agreement with the experimental binding energies.
For $U=7.5$\,eV (purple) the $4f$ band lies at the top of Au substrate $d$ band.}
\label{fig:filling_reference}
\end{figure}

\clearpage

\bibliography{maere}

\makeatletter\@input{aux4_EuAu2_GdAu2_250122.tex}\makeatother